\documentclass[10pt,sigconf]{acmart}

\acmConference[ ]{ }

\setcopyright{none}
\renewcommand\footnotetextcopyrightpermission[1]{}
\newcommand{\smartparagraph}[1]{\noindent{\bf #1}\ }

\title{Leveraging Large Language Models to Contextualize Network Measurements}

\author{Roman (Sylee) Beltiukov}
\affiliation{%
  \institution{UC Santa Barbara, USA}
  \country{}
}

\author{Karthik Bhattaram}
\affiliation{%
  \institution{UC Santa Barbara, USA}
  \country{}
}

\author{Evania Cheng}
\affiliation{%
  \institution{UC Santa Barbara, USA}
  \country{}
}

\author{Vinod Kanigicherla}
\affiliation{%
  \institution{UC Santa Barbara, USA}
  \country{}
}

\author{Akul Singh}
\affiliation{%
  \institution{UC Santa Barbara, USA}
  \country{}
}

\author{Ken Thampiratwong}
\affiliation{%
  \institution{UC Santa Barbara, USA}
  \country{}
}

\author{Arpit Gupta}
\affiliation{%
  \institution{UC Santa Barbara, USA}
  \country{}
}

\begin{document}

\maketitle

With the worldwide growth of remote communication and telepresence~\cite{trafficgrowth}, network measurements form a cornerstone of effective performance assessment and diagnostics for Internet users. The value of such measurements is particularly evident when pinpointing network bottlenecks or validating service-level agreements with the Internet Service Provider. Most often, users seek for overall connection performance measurement using publicly available tools (also known as `speed tests'~\cite{ookla, mlab, cloudflare}) that provide an overview of their connection's throughput and latency.

However, extracting meaningful insights from these measurements remains a challenging task for a non-technical audience. Interpreting network measurement data often requires considerable domain expertise to account not only for subtle variations of the connection stability and metrics, but even for simpler concepts such as latency under load or packet loss influence towards connection performance. In the absence of proper expertise, common misconceptions can easily arise. A notable example is the widespread reliance on download bandwidth measurements~\cite{10.1145/3517745.3561441} reported by speed test tools, which might give a false sense that the connection achieves the expected performance. In reality, the connection stability, packet loss, latency variations, and other factors can significantly impact the test interpretation, introducing disconnection between reported test results and subjective user experience, or leading to misguided decision-making when provisioning or troubleshooting network services.

To address these issues, researchers should recognize the importance of making network measurements not only more comprehensive but also more accessible for wider audience without deep technical knowledge. A promising direction to achieve this goal involves leveraging recent advancements in large language models (LLMs), which have demonstrated capabilities in conducting an analysis of complex data in other fields, such as laboratory test results interpretation~\cite{info:doi/10.2196/56655}, news summarization~\cite{10.1162/tacl_a_00632}, and personal assistance~\cite{10.1145/3580305.3599572}.

In this paper, we describe an ongoing effort to apply large language models and historical data to enhance the interpretation of network measurements in real-world environments. We aim to automate the translation of low-level metric data into accessible explanations, allowing non-experts to make more informed decisions regarding network performance and reliability.



\begin{figure}
    \centering
    \includegraphics[width=0.8\linewidth]{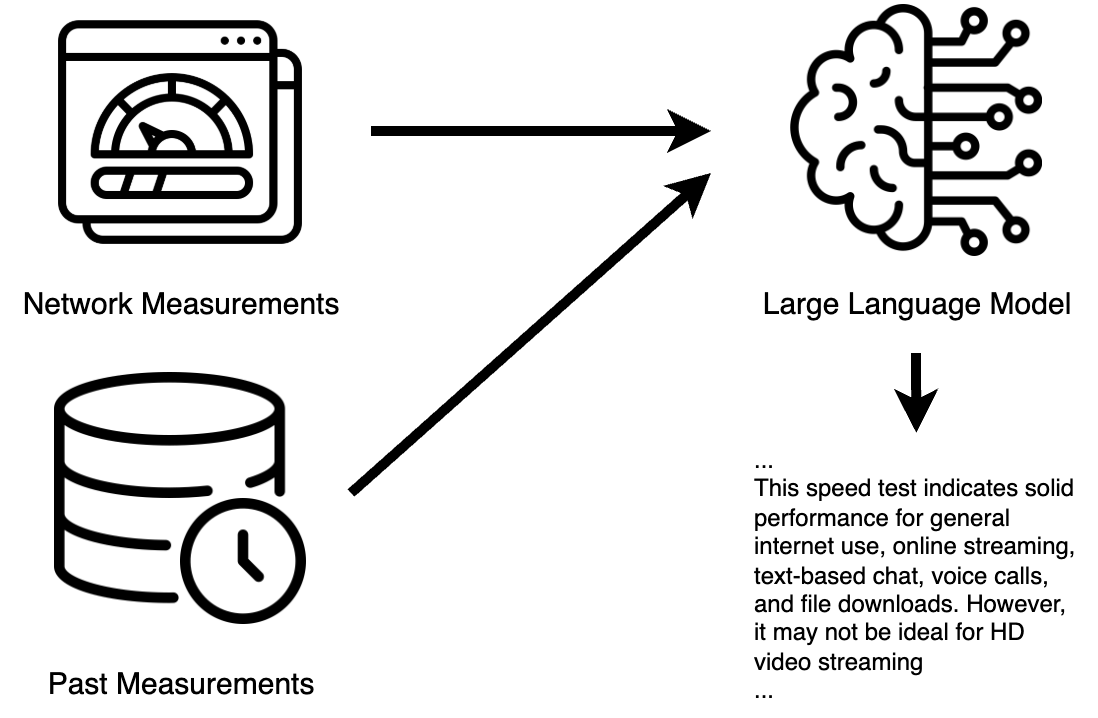}
    \caption{Components of the network measurements framework.}
    \label{fig:scheme}
\end{figure}

\section{Proof of Concept}
To demonstrate the potential of LLMs in enhancing network measurement interpretation, we adapted a publicly available framework for network measurements, netUnicorn~\cite{netunicorn}, to incorporate Large Language Models (LLMs) for parsing and interpreting the results. We developed a custom netUnicorn pipeline that allows users to run network measurements (speedtest) using the Ookla CLI tool~\cite{ookla}. The results contain widely used metrics (such as download and upload throughput, latency, etc) which are often presented to the user in the raw format. To improve understanding of these results, we incorporated an additional step in the pipeline that sends the raw results to a separate endpoint for further interpretation and analysis. This endpoint filters the results, removing unnecessary or private information, i.e., IP and MAC addresses, test ID, timestamp, unique URLs, etc. 

After cleaning up the raw results, we apply a Large Language Model to interpret and write a summary for the user. For this purpose, we used a locally hosted version of the DeepSeek-R1 Distill Qwen model with 7B parameters, trained on a diverse corpus of text data, including technical documentation and websites content. The model was hosted using LM Studio~\cite{lmstudio}, which provides an endpoint for text analysis and generation requests using an OpenAI API-compatible interface. The choice of the interface and the hosting solution allows us to easily integrate different models (including open-source and proprietary) into our pipeline, providing flexibility and scalability for future improvements.

To improve the accuracy of the results, we also added a geolocation step to the pipeline. This step uses the user's IP address to determine their location via IP Geolocation API~\cite{ipapi} and then provides approximate anonymized information (with city-level granularity) about the user's location to the LLM. This information is used to provide more relevant and localized insights into the network performance, taking into account the user's geographical context and associated with it network conditions.

To further enhance the fidelity of network measurements interpretation, we explored the usage of Retrieval-Augmented Generation (RAG) techniques which combine the strengths of LLMs with external knowledge sources containing historical data, allowing the model to retrieve relevant information from a database or knowledge base and incorporate it into the model responses. For our framework, we implemented a knowledge database of past network measurements using publicly available data from the FCC Measuring Broadband America program~\cite{fccmba}. This data contains various measurements from different locations and ISPs, including download and upload throughput obtained via running speed test software. After receiving the raw measurement results from the netUnicorn pipeline, our framework queries the local PostgreSQL knowledge database to find similar measurements based on the measurement server location and time of day. We calculate statistics over retrieved data, such as average fetch time, throughput mean, median, and standard deviation values, and provide them to the LLM as an additional context together with the raw results and geolocation information to use the historical context of their measurements to provide more accurate and relevant insights into the current network performance. 

We designed and implemented a custom prompt for the LLM that includes instructions to focus on key metrics while also considering the contextual information (i.e., key spatial, temporal, and semantic relationships between measurements from different tools, locations, and times). This guides the model to provide a concise and informative summary that highlights the most important aspects of the results, targeting a non-technical audience to facilitate the analysis.

The resulting summary is sent back to the user, providing them with overall information, a detailed explanation of each metric, and their potential impact on user experience for different use cases, such as online gaming, video streaming, internet browsing, etc. The summary also might include recommendations for improving network performance if applicable, such as optimizing wireless access point placement, using wired connections, or upgrading the internet plan.

\section{Further research}

The current implementation of our framework has several significant limitations. First, the choice of large language models can significantly impact the generated summaries and bigger models may provide more accurate results and prevent hallucinations or misinterpretations by increasing the cost of running the framework and require more computational resources (for both inference and fine-tuning the model on specific networking literature for analysis improvement). Second, the accuracy of the geolocation step depends on the quality of the IP address lookup services and methods the user or ISPs are using to hide their current location. Finally, the current knowledge database is limited to the FCC MBA program data, which may not provide all desired metrics or cover all geographical areas. 

More fundamentally, we aim to explore challenges that arise when using LLMs for measurements interpretation:

\smartparagraph{Trust and reliability.} How can we ensure that the LLM-generated summaries are accurate and trustworthy? What measures can be taken to validate the results and prevent misinformation or hallucinations of the model?

\smartparagraph{Overemphasis on certain metrics or contextual information.} Can we prevent the model from highlighting certain metrics or contextual information that may not be relevant to the user's specific situation? How can we ensure that the model provides a balanced and comprehensive analysis of the network performance?

\smartparagraph{Data diversity and bias.} Given that performance of LLM is shaped by the training data, how can we make results less biased and more representative of the real-world network?%

\smartparagraph{Privacy issues.} How can we strike a balance between providing relevant contextual information and user privacy? 

\smartparagraph{User experience.} Can we ensure that provided results significantly improve the user experience and understanding of network performance? How can we design the framework to be user-friendly and accessible to a non-technical audience?

We believe that addressing these questions and challenges is crucial for the reliable and useful implementation of LLMs in network measurements interpretation, which will avoid misleading or incorrect conclusions and provide users with accurate and actionable insights into their network performance.

\bibliographystyle{abbrv}
\bibliography{references}

\end{document}